\documentclass{article}[12pt]
\usepackage{amsmath}
\usepackage{amssymb}
\usepackage{amsthm}
\usepackage{amsfonts}           
\usepackage{rsfs}
\usepackage[dvips]{graphicx}

\def\abstract#1{\vskip 7mm
        \begin{center}{\large Abstract}\par \smallskip
                \begin{minipage}[c]{12cm}
                        \small #1
                \end{minipage}
        \end{center}
}

\def\title#1{\begin{center}{\Large\bf #1}\end{center}}
\def\author#1{\vskip 5mm \begin{center}{#1}\end{center}}
\def\address#1{\begin{center}{\it #1}\end{center}}
\newcommand{\ssmatrix}[4]%
{\begin{pmatrix} #1 & #2 \\ #3 & #4 \end{pmatrix}}
\makeatletter

\@ifundefined{lesssim}{}{}
\@ifundefined{gtrsim}{}{}
\def\vereq#1#2{\lower3pt\vbox{\baselineskip1.5pt \lineskip1.5pt
\ialign{$\m@th#1\hfill##\hfil$\crcr#2\crcr\sim\crcr}}}
\makeatother

\newtheorem{theorem}{Theorem}
\newtheorem{definition}{Definition}
\newtheorem{corollary}{Corollary}
\newtheorem{remark}{Remark}
\newtheorem{lemma}{Lemma}
\newtheorem{proposition}{Proposition}

\newcommand{\UU}{{\mathscr U}}
\newcommand{\OO}{{\mathscr O}}

\begin{document}

\title{%
Half period theorem of binary black holes}
\author{%
  Masaru Siino\footnote{E-mail:msiino@th.phys.titech.ac.jp} 
and Daisuke Ida 
}
\address{%
  Department of Physics, Tokyo Institute of Technology, \\
  Oh-Okayama 1-12-1, Megro-ku, Tokyo 152-8550, Japan \\
  Department of Physics, Gakushuin University, Tokyo 171-8588, Japan
}

\abstract{
Merging event horizons of the binary black holes is investigated.
While recent development of the numerical study of the binary black hole  coalescence has shown that their apparent horizons can
orbit for many periods, we study the orbital motion of the event horizon.
We
discuss how many periods their event 
horizons orbit before their coalescence.
Then, we find that they soon merge into one and the black holes cannot 
orbit for a half period while the apparent horizons can orbit many times.
}

%%%%%%%%%%%%%%%%%%%%%%%%%%%%%%%%%%%%%%%%%%%%%%%%%%%%%%%%%%%%%%%%%%
\section{Introduction}
Recently the black hole binary system is much paid attention to as a candidate for gravitational wave source.
Indeed numerical studies of the binary black hole coalescence advances  
 so that
it predicts the time profile of the gravitational radiation.
Usually in numerical simulation, the black hole formation is examined by determining an apparent horizon. 
This is because the existence of the apparent horizon strongly suggests that of the event horizon. The event horizon is meaningful for an asymptotic observer, since he can just observe the outside of the event horizon, but not the apparent horizon. 
Moreover the event horizon is gauge invariant concept, while apparent horizon, even its existence depends on timeslicing.
Hence, we would like to discuss the event horizon for the binary black hole system.

Some authors have made sure that two apparent horizons can orbit around each other in a case of the  binary black hole coalescence\cite{NS1}\cite{NS2}.
Then we attempt to discuss whether this picture also holds for the event horizon.
Reason why we suspect the behavior of event horizon differs from that of the apparent horizon is that, in some studies of black hole formation, it is insisted
that the event horizon soon settles to a single sphere\cite{Vank}\cite{MS1}.
In other words, we speculate that there does not remain sufficient time for the two black holes to orbit around each other.

\begin{figure}[hbtp]
\centering
\includegraphics[width=10cm,clip]{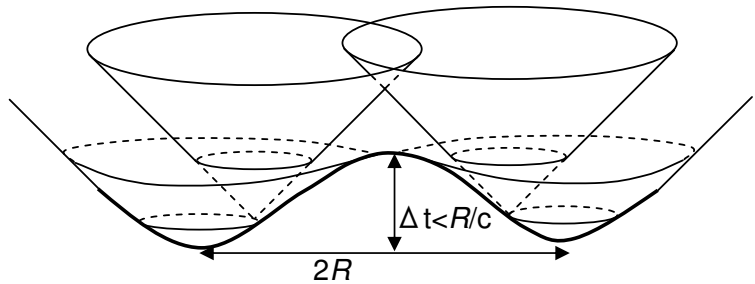}
\caption{An event horizon and its crease set. The crease set is achronal and roughly implies two event horizon coalesce in superluminal speed.}
\label{fig:figi}
\end{figure}

It is known that
 the event horizon is generated by null geodesics without a future endpoint~\cite{HE}.
In particular, the event horizon is a null hypersurface which may not be smooth at the past endpoints of the null geodesic generators. The set of the past endpoints of the null geodesic generators of the event horizon will be called the {\it  crease set}~\cite{BK}\cite{MS1}. 
Since the crease set is a subset of the null hypersurface, it is an achronal set(any two points are not connected by a timelike curve) as illustrated in Fig.\ref{fig:figi}~\cite{Win}\cite{MS1}. That the crease set is achronal roughly implies two event horizons coalesce in superluminal speed.

As a rough estimation, let two black holes be located with separation $2R$.
They will coalesce within the time-interval of order of $\Delta t=R/c$ while we cannot expect any circular motion with this time duration.
Hence we expect there is any upper bound of a rotation angle of the binary black holes before their coalescence.

In the present article, we assert the existence of the upper bound of the rotation angle 
and demonstrate a half period theorem that binary black holes cannot orbit for a half period 
in  terms of their event horizons 
assuming the reflection symmetry with respect to the orbital surface.
This assumption is required for technical reason 
to formulate the notion of the half period of the binary black holes without 
ambiguities.
Clearly, it is meaningless to say that the binary black holes go around or half around
in terms of a coordinate system.
We should formulate the half period of the binary black holes without referring to a
specific coordinate system.

The outline of our argument is as follows.
We first consider the foliation of the spacetime by a family of timelike curves 
and choose a reference timelike curve as a center of the binary motion.
Next we determine the opposite direction,
which we call the {\em light ray opposite},
of each black hole with respect to
the center. Then, whether the black hole orbits for a half period from the
initial configuration or not can be determined in terms of the light ray opposite.
Finally, we show that each black hole cannot orbit for a half period irrespective of
the choice of the foliation and the center.
This gives a possible way to measure the amount of the binary motion 
in a coordinate-free way.

We first 
explain more details of our scheme in the next section.
In the third section we 
give several definitions and set up the situation.
The main theorem is 
stated 
in the fourth section.
In the last section, we give some remarks and implications.
%%%%%%%%%%%%%%%%%%%%%%%%%%%%%%%%%%%%%%%%%%%%%%%%%%%%%%%%%%%%%%%%%%%%%
\section{schematic picture}
Now, we try schematic discussion based on the topological notion of event horizon to illustrate that a half period of the binary rotation typically bounds   duration of binary rotating era.
From Newtonian picture, one may think that the coalescence of binary black hole event horizons occurs
after two black hole event horizons orbit around each other for several periods.
A simple picture of the orbiting event horizons, however, makes us suspect them of merging before the several periods have elapsed.

\begin{figure}[hbtp]
\centering
\includegraphics[width=14cm,clip]{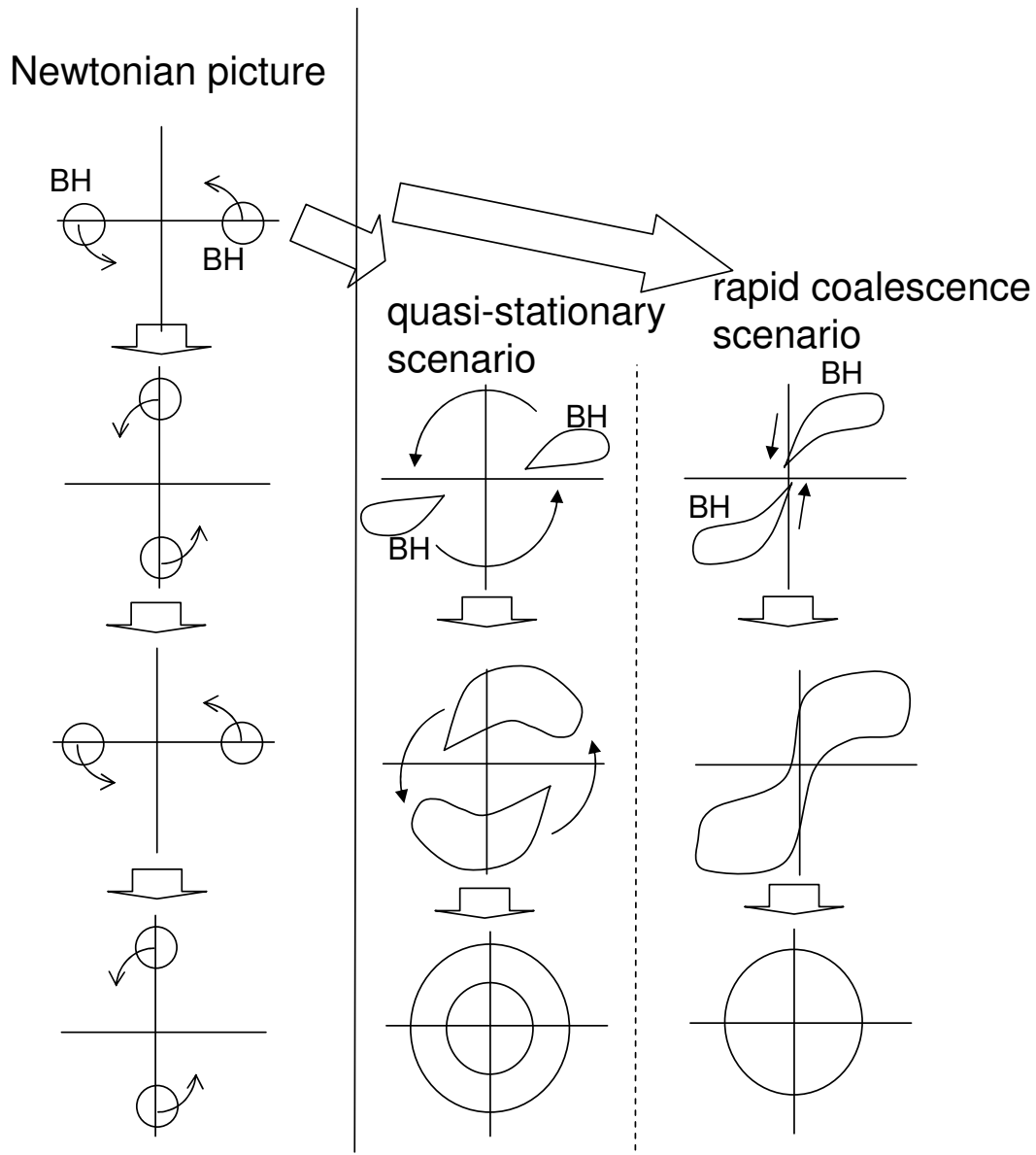}
\caption{The figures in the left side illustrate a Newtonian picture of binary motion. The others are two possible scenarios for binary black hole coalescence. The center is the quasi-stationary scenario and the right is the rapid coalescence scenario.}
\label{fig:fig0}
\end{figure}

Here we simply consider two 
black holes with identical mass
fated to coalesce. 
They are on a binary orbit shrinking by the energy loss caused by gravitational radiation.
In a coordinate system $(t,x^i)$
such that each $x^i={\rm const.}$ line is timelike,
every comoving observer inside a black hole at some moment
will stay 
 within the black hole region,
while the black hole move on the binary orbit.
This implies that a Newtonian picture of binary motion as 
in Fig.\ref{fig:fig0}
is incorrect.

We expect two possible scenarios for binary black hole coalescence,
which we call the quasi-stationary scenario and the rapid coalescence scenario,
 as follows.
The quasi-stationary scenario is what we expect to occur in the quasi-stationary binary motion
of black holes.
In the quasi-stationary scenario, the event horizons form a toroidal event horizon
along the binary orbit in a half period of binary motion (See Fig.\ref{fig:fig0}).
The rapid coalescence scenario is expected to occur when the binary black holes don't have 
orbital angular momentum enough.
In this case, two black holes coalesce without forming a toroidal horizon (See Fig.\ref{fig:fig0}).
In both scenarios, binary black holes do not orbit for a half period before coalescence.

In the above argument, the term ``a half period'' is loosely used.
It is in general difficult to determine whether the binary black holes go around
without specific coordinate system.
The present work is our attempt to
seek for a rigorous statement corresponding to the above schematic discussion.
For this purpose,
we should study it in a generally covariant manner 
incorporating a technique based on the causal studies of general relativity.

\section{set up}
First of all, we have to consider the amount of orbital rotation of the binary black holes.
This will be determined by introducing a global angular coordinate function.
However, it is in general difficult due to the lack of the standard coordinate system.

For this reason, we give up considering one radian to be an upper bound of 
orbital
rotation angle, and attempt to
put a mark corresponding to a half period in order to check 
orbiting of black holes.
We ought to be able at least to discuss a half period by introducing a `straight' curve
passing the antipodal point, even without the angular coordinate.

Nevertheless, each black hole does not always pass the opposite side of 
the straight curve in general cases.
For simplicity in the present work, we assume reflection symmetry such that 
the plane of symmetry, which we call the orbital surface, intersects binary black holes.
With the reflection symmetry, it would be enough to discuss the motion of 
the object only on the orbital surface. 

We define a {\it light ray opposite} as the mark of
 a half of the orbital period.
Suppose a spacetime $(M,g)$ is globally hyperbolic.
Then, $(M,g)$ admits timeslicing $\{\Sigma(t)\}$ 
in terms of a global time function $t:M\to {\bf R}$
and
there will be a timelike vector field $T$ without zero points.
Let the vector field $T$ be normalized such that $\langle T,dt\rangle=1$ holds.

\begin{figure}[hbtp]
\centering
\includegraphics[height=6cm,clip]{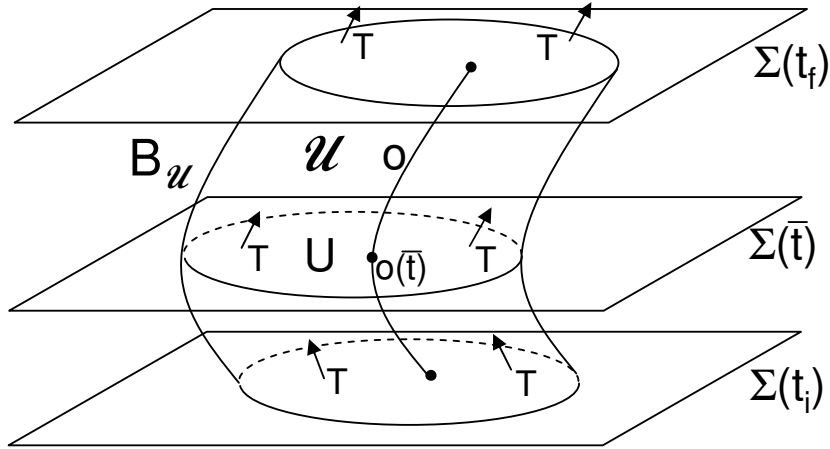}
\caption{A cylinder defined by the 4-tuple $(\Sigma(t),\UU,T,o(t))$ is the 
comoving ball with the origin.}\label{fig:fig1}
\end{figure}

Let us consider a cylindrical region $\UU$ generated by the vector field $T$
with a timelike side-boundary $B_{\UU}$ as illustrated in figure \ref{fig:fig1}.
Then the timelike vector field $T$ 
determines a natural projection $$\pi_t:\UU\mapsto \UU\cap\Sigma(t)$$
of the closed subset $\UU$ of $M$ into each
timeslice $\Sigma(t)$ along the integrated curve of $T$.

\begin{definition}
[Comoving ball with the origin]
Let $(M,g)$ be a globally hyperbolic space-time.
Let $t: M\to {\bf R}$ be a global time function in $M$.
The sequence $\{\Sigma(t)\}$ of the $t={\rm const.}$ hypersurface  $\Sigma(t)$, $t\in [t_i,t_f]$ 
will be called the timeslicing.
The space-time $M$ will admit a future directed timelike vector field $T$ 
such that $\langle T, dt\rangle=1$ holds.
Let $U\subset\Sigma(\overline{t})$ be a 
topological 3-ball embedded in %compact subset of 
 $\Sigma(\overline{t})$, $\overline{t}\in(t_i,t_f)$. %for some fixed $\overline{t}$, $(t_i<\overline{t}<t_f)$
Let $\UU$ be the closed subset of $M$ generated by $T$ such that 
$\UU\cap\Sigma(\overline{t})=U$ holds.
Let $B_{\UU}\subset \dot\UU$ be the closed subset of $\dot\UU$, which is the 
product set $\dot U\times [t_i,t_f]$ generated by $T$, 
and let 
$o: [t_i,t_f]\to M; t\mapsto o(t)$ %$o(t)$ 
be the integral curve of $T$ which pass through 
an interior point $o(\overline{t})$ of $U$.

The 4-tuple $(\{\Sigma(t)\},\UU,T,o)$ will be called the comoving ball with the origin 
in $(M,g)$.
\qed
\end{definition}

It is difficult to give a precise notion of orbital plane of binary black holes
or period of binary motion in general settings.
A possible way to overcome this difficulty is to impose a reflection symmetry 
of $(M,g)$. 
Although this restriction might be rather stringent, 
we could still consider a large class of space-times describing binary black holes.
In the rest of the present paper, we will assume that the space-time $(M,g)$ admits 
a reflection symmetry with respect to a timelike hypersurface $\OO$ in $M$.
The fixed point set $\OO$ under this isometry will be called the orbital surface
and the orbital surface at the time $t$ will be denoted as
$\OO_t=\OO\cap \Sigma(t)$.
All the settings including the global time function $t$ and
the comoving ball with the origin $(\{\Sigma(t)\},\UU,T,o)$ are taken as respecting the
reflection symmetry.

\begin{figure}[hbtp]
\centering
\includegraphics[width=9cm,clip]{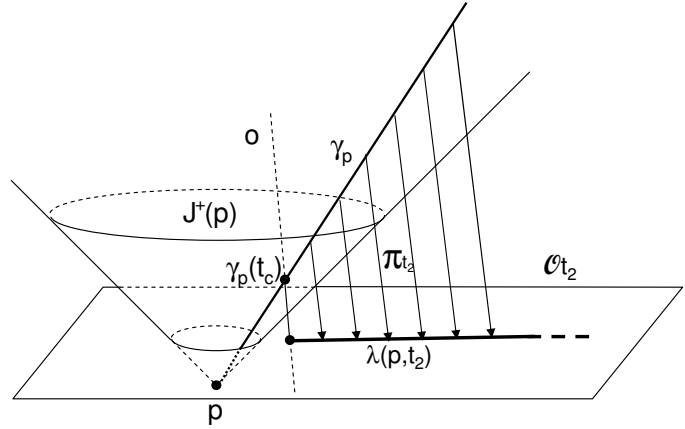}
\caption{A light ray opposite $\lambda$ on $\Sigma(t_2)$ with respect to $p$ is given by $T$-projection of $\gamma(t)$.}\label{fig:fig2}
\end{figure}

Here we introduce a notion of the opposite side of a point $p$ beyond the origin $o$,
which we call the light ray opposite (abbreviated to LRO) of $p$,
in terms of a null geodesic generator of $J^+(p)$ (see Fig.~\ref{fig:fig2}). 

\begin{definition}[The light ray opposite]
Let $(M,g)$ be a reflection symmetric spacetime with respect to the
timelike hypersurface $\OO$.
Let $(\{\Sigma(t)\},\UU,T,o)$ be a comoving ball with the origin in $M$
respecting the reflection symmetry.
Let us call $\OO_t=\OO\cap \Sigma(t)$ the orbital surface at the time $t$.
For a point $p\in\OO_{t_1}\cap\UU$, $t_1\in(t_i,t_f)$,
$\gamma_p$ 
is defined to be the geodesic generator of $\dot J^+(p)$ from $p$ 
which pass through $o$ at the time $t=t_c\in(t_1,t_f)$,
if there is exactly one null geodesic generator of $\dot J^+(p)$ from $p$ through $o$.
The null geodesic $\gamma_p$ will be within $\OO$ due to the reflection symmetry.
Then, for a point $p$
and  a time $t_2\in (t_1,t_f)$,
the light ray opposite  (LRO)
$\lambda(p,t_2)$ of $p$
at the time $t=t_2$ is 
defined by
\begin{align*}
\lambda(p,t_2)=\pi_{t_2}\left[\gamma_p\cap \bigcup_{t\in[t_c,t_f)}\Sigma(t)\right]
\end{align*}
where $\pi_{t_2}$ denotes the projection $\UU\to \Sigma(t_2)$
naturally defined by the timelike vector field $T$, and 
the LRO of $p$ is defined to be the empty set if $\gamma_p$ is not defined.

\qed
\end{definition}

\begin{remark}
The LRO is the empty set when the origin is far from the reference point $p$
such that $o~\cap J^+(p)=\emptyset$.
Besides there might possibly occur the situation where $o~\cap J^+(p)\ne \emptyset$ holds
but $o~\cap \dot J^+(p)=\emptyset$, when the congruence of the light rays from $p$
to $o$ have caustics due to the local gravitational effects.
We have just precluded such a possibility for simplicity.
\end{remark}

\begin{figure}[hbtp]
\centering
\includegraphics[width=9cm,clip]{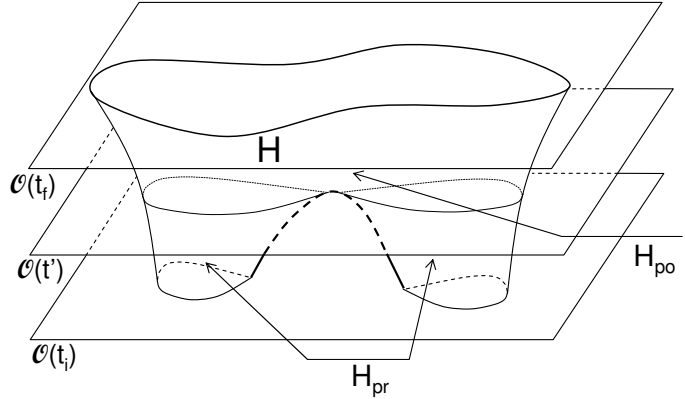}
\caption{This is a binary black hole coalescence system which is composed of a disconnected pre-coalescence part and a connected post-coalescence part.}\label{fig:fig3}
\end{figure}

Next, we introduce the notion of coalescing binary black holes as follows
(Fig.~\ref{fig:fig3}).

\begin{definition}[The binary black hole coalescence system]
Let $(M,g)$ be a black hole spacetime 
and let $(\{\Sigma(t)\},\UU,T, o(t))$ be a comoving ball with the origin in $M$.
A binary black hole coalescence system $(H,\{\Sigma(t)\})$ is 
defined to be
the pair consiting of %a couple of 
the event horizon $H$ in $M$
and
a timeslicing $\{\Sigma(t)\}$, $t\in [t_i,t_f]$,
such that there is a coalescence time $t'\in(t_i,t_f)$
decomposing $H\cap\UU$  into a pre-coalescence part 
$$H_{pr}=H\cap \UU\cap \left[ \bigcup_{t\in[t_i,t')}\Sigma(t)\right]$$
which 
has a pair of connected components,
and a post-coalescence part 
$$H_{po}=H\cap \UU\cap \left[ \bigcup_{t\in (t',t_f]}\Sigma(t)\right]$$
which is connected,
 by the spatial hypersurface $\Sigma(t')$.
\qed
\end{definition}

%Here it 
It
should be noted that the concept of the black hole coalescence 
depends on the choice of %is not independent of 
the timeslicing.
In a different timeslicing
the black hole coalescence system can 
always be
regarded as the formation of a single black hole~\cite{MS1,MS2}. 
Here we consider the typical case $S^2\sqcup S^2\to S^2$ for the transition 
of the horizon topology.

Since the reflection symmetry of $M$ is imposed, 
it is enough to consider the causal structure of the orbital surface
$\OO$.
In other words, we will concentrate on the section of the event horizon by $\OO$.
In the following, we assume that $H_{pr}\cap\OO_t$
consist of a pair of circles $S^1\sqcup S^1$
and $H_{po}\cap\OO_t$ consists of a single circle $S^1$
in the binary black hole coalescence system.

We want to %formulate 
define
the half cycle of the binary black hole system 
by saying that %in such a way 
that
a black hole go around a half cycle 
if %by saying that 
all the infinitesimal
area elements of the black hole go around a half.
For this purpose, we need to 
specify each trajectory of a point on the black hole. 
At first sight, 
the null geodesic generators of the event horizon seems to naturally determine each orbit.
However, 
this is not appropriate, since new null geodesic generators emerge incessantly.
Instead, we consider arbitrarily chosen 
one parameter family of homeomorphisms, %one parameter family of homeomorphisms,
$\phi_t: S^1\to S^1$
between the $H_{pr}\cap \OO_{t_i}$ and $H_{pr}\cap \OO_{t}$, $(t\in [t_i,t'))$,
continuous with respect to $t$.
Such one parameter family of homeomorphisms $\phi_t$ %naturally 
determines
the motion of the infinitesimal area element of the black hole.
Note that each orbit determined by $\phi_t$ necessarily exceeds or equals the speed of light, 
for it lies on the null hypersurface.

\begin{figure}[hbtp]
\centering
\includegraphics[width=9cm,clip]{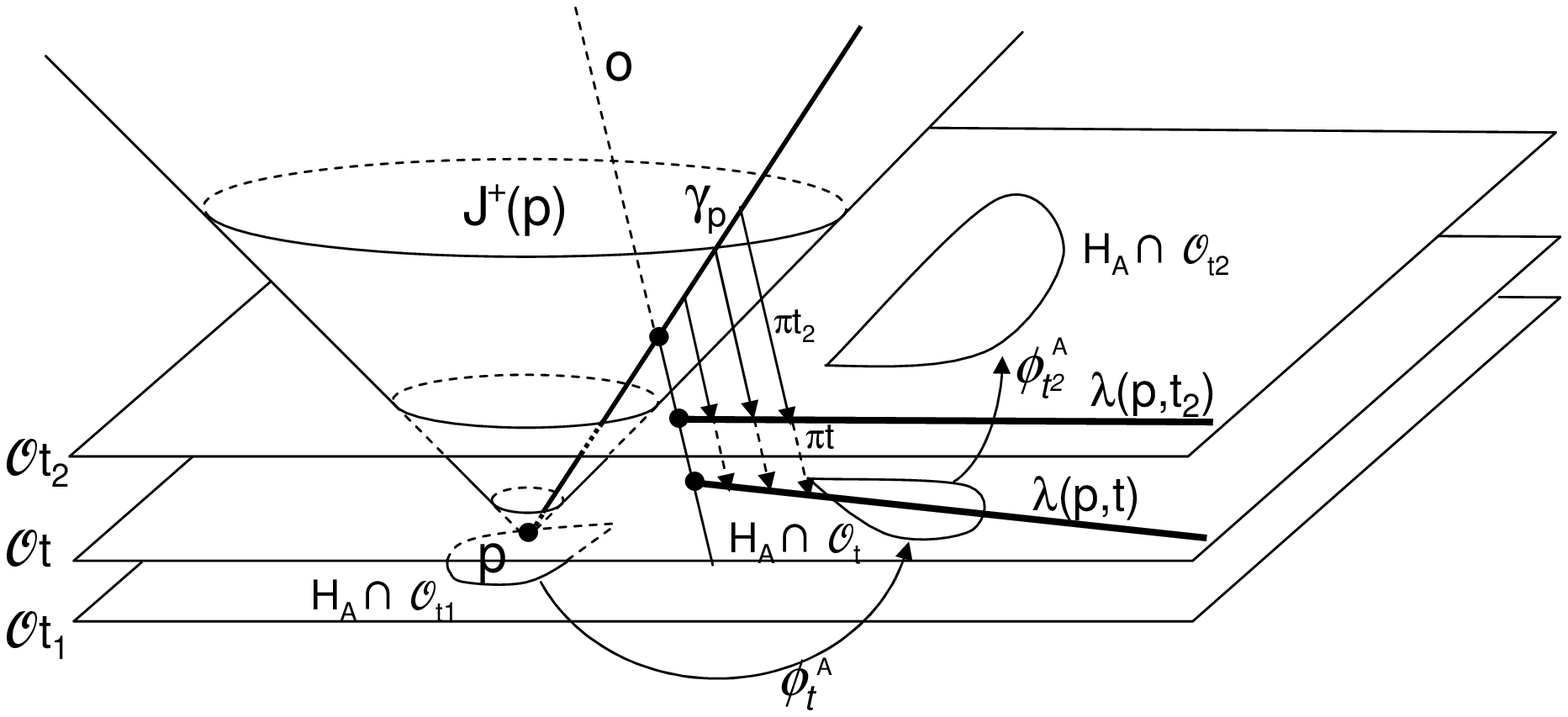}
\caption{The definition of a half period of 
the binary coalescence system.} %binary black holes in an orbital surface.}
\label{fig:fig4}
\end{figure}

The following is the definition of a half period of binary black holes in an orbital surface, where each black hole event horizon moves as shown in figure\ref{fig:fig4}.
\begin{definition}[A half period of 
the binary coalescence system] %binary black holes  in the orbital surface]
\label{def:half}
Let $(H,\{\Sigma(t)\})$ be a binary black hole coalescence system with the reflection symmetry 
with respect to the orbital surface $\OO$ %${\OO_t\subset \Sigma(t)}$ for $t\in (t_i,t_f)$.
with the coalescence time $t'\in (t_i,t_f)$.
Let the pre-coalescence part $H_{pr}\cup \OO$ in $\OO$
consists of %disconnected sum,
a disconnected sum
$H_{pr}\cup \OO=H_{\rm I}\sqcup H_{\rm II}$
such that each of $H_{\rm I}\cup  \OO_t$ and $H_{\rm II}\cup \OO_t$, $t\in [t_i,t')$
is a circle.
Let 
$\phi_t^A: H_A\cap \OO_{t_1}\to H_A\cap\OO_t$ ($A={\rm I}$, ${\rm II}$)
be a one parameter family of homeomorphisms such that $\phi_{t_1}^A$ is the identity
map of $H_A$
and $\phi_t^A$ is continuous with respect to $t$.
%For a comoving ball with the origin, we 
For $t_i<t_1<t_2<t_f$,
we say that a half period of 
the binary coalescence system %binary black holes 
has elapsed during $(t_1,t_2)$, %$t_1$ and $t_2\in (t_1,t')$, 
if the following statement hold both for $A={\rm I}$ and ${\rm II}$;
There is a point $p$ on $H_A\cap O_{t_1}$ such that
every  orbit %All the orbits
of a point on
 $H_A\cup \OO_{t_1}$ intersects the LRO $\lambda(p,t)$ of $p$ during  $t_1<t<t_2$.
\qed
\end{definition}

Note that
%According to this definition,
whether the half period 
has %have 
elapsed or not does not depend on the choice of the correspondences
 $\{\phi^{\rm I}_t,\phi^{\rm II}_t\}$. 

%%%%%%%%%%%%%%%%%%%%%%%%%%%%%%%%%%%%%%%
\section{The half period theorem}

\begin{figure}[hbtp]
\centering
\includegraphics[width=9cm,clip]{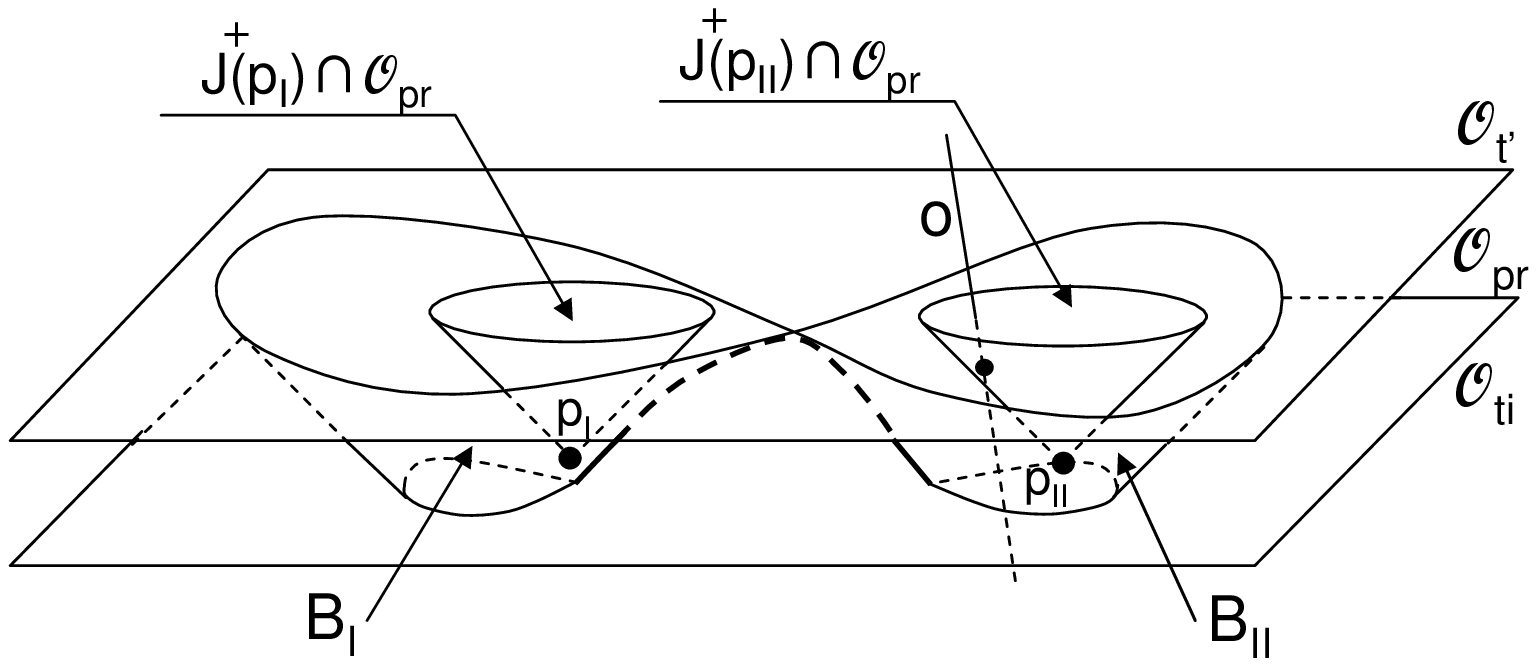}
\caption{Two black holes are causally separated. }\label{fig:fig5}
\end{figure}

Now we show that a half period of 
the binary coalescence system %binary black holes 
does not
elapse before the coalescence. % elapses.
Firstly,  it is easily seen %we see 
that two black holes before the coalescence are causally separated 
with each other %
as illustrated in figure~\ref{fig:fig5}.  
\begin{proposition}
Let ${\mathscr B}$ be the black hole region in 
the binary black hole system with reflection symmetry, 
and let $\OO_{pr}$ be the part of $\OO\cap \UU$ before the coalescence time defined by
$\OO_{pr}=\bigcup_{t\in[t_i,t')}\OO(t)\cap \UU$, %$\UU_{pr}=\UU\cap\{\Sigma(t)|t<t'\}$,
so that the black hole region $B={\mathscr B}\cap \OO_{pr}$ in the orbital surface
is composed of 
two 
black hole regions $B_{\rm I}\simeq D^2\times [t_i,t')$ 
and $B_{\rm II}\simeq D^2\times [t_i,t')$
without intersection, where $D^2$ denotes the closed 2-disk.
Then,
\begin{align*}
\forall p_{\rm I}\in B_{\rm I}, 
\forall p_{\rm II}\in B_{\rm II},\quad
\OO_{pr}\cap J^+(p_{\rm I})\cap J^+(p_{\rm II})=\emptyset
\end{align*}
holds.
\label{prop:1}
\end{proposition}
\begin{proof}
Since $p_A$ ($A={\rm I}$, ${\rm II}$) is a point in $B_A$,
and $B_A$ is a future set in $\OO_{pr}$, $J^+(p_A)\cap\OO_{pr}\subset B_A$ holds.
%Therefore, $J^+(p_{\rm I})\cap \UU_{pr}$ and $J^+(p_{\rm II})\cap \UU_{pr}$ 
%do not intersect, for $\BB_a\cap\BB_b=\emptyset$.
Then, $B_{\rm I}\cap B_{\rm II}=\emptyset$ implies
$(J^+(p_{\rm I})\cap \OO_{pr})\cap (J^+(p_{\rm II})\cap \OO_{pr})=\emptyset$.
\end{proof}

This proposition just reflects the fact that 
a black hole region is a future set. 
The following corollary 
immediately follows. %Consequently, we easily see the following corollary. 
\begin{corollary}
A timelike curve $o$ in $\OO$ does not intersect
both with $J^+(p_{\rm I})\cap\OO_{pr}$ and $J^+(p_{\rm II})\cap\OO_{pr}$
for any $p_{\rm I}\in B_{\rm I}$ and $p_{\rm II}\in B_{\rm II}$.
\label{col:1}
\end{corollary}

Next, we show the following lemma.
\begin{lemma}
For every pair of points
$p_{\rm I}\in B_{\rm I}\cap \OO_{t_1}$ and $p_{\rm II}\in B_{\rm II}\cap\OO_{t_1}$, 
either of the following statements holds.
\begin{enumerate}
\item The LRO $\lambda(p_{\rm I},t_2)$ of $p_{\rm I}$ does not intersect 
with %is not contained in 
$J^+(p_{\rm I})$ for any $t_2\in(t_1,t')$.
\item The LRO $\lambda(p_{\rm II},t_2)$ of $p_{\rm II}$ does not intersect with %is not contained in 
$J^+(p_{\rm II})$ for any $t_2\in(t_1,t')$.
\end{enumerate}
\label{lem:1}
\end{lemma}
\begin{proof}
It follows from the Cor.~\ref{col:1}
that, for $p_{\rm I}\in B_{\rm I}\cap\OO_{t_1}$ and $p_{\rm II}\in B_{\rm II}\cap\OO_{t_1}$, 
$o$ does not intersect both with $J^+(p_{\rm I})\cap\OO_{t_1}$ and
 $J^+(p_{\rm II})\cap\OO_{t_1}$. Assume $J^+(p_{\rm I})\cap\OO_{t_1}$ does not intersect $o$.
Let the LRO $\lambda(p_{\rm I},t_2)$ have an intersection with $J^+(p_{\rm I})\cap\OO_{pr}$,
then $\lambda(p_{\rm I},t_2)$ starts from $o(t_2)$ and extends to the point $q$ on $\dot J^+(p_{\rm I})
\cap \OO_{pr}$. It follows from the definition of the LRO that $q$ is on the past directed
timelike curve generated by $T$ 
starting from the point $p$ on $\dot J^+(p_{\rm I})$. 
By slightly deforming the causal curve from $p_{\rm I}$ to $p$ 
obtained by joining
the null geodesic generator from $p_{\rm I}$ to $q$ and the timelike curve from $q$ to $p$,
one can construct the timelike curve from $p_{\rm I}$ to $p$.
It follows that there is an open neighborhood $U$ of $p$, such that
$U$ is contained in the chronological future $I^+(p_{\rm I})$ of $p_{\rm I}$.
This contradicts the fact that the neighborhood $U$
of the boundary point $p$ of $J^+(p_{\rm I})$
necessarily contains an exterior point of $J^+(p_{\rm I})$.
\end{proof}

\begin{figure}[hbtp]
\centering
\includegraphics[width=9cm,clip]{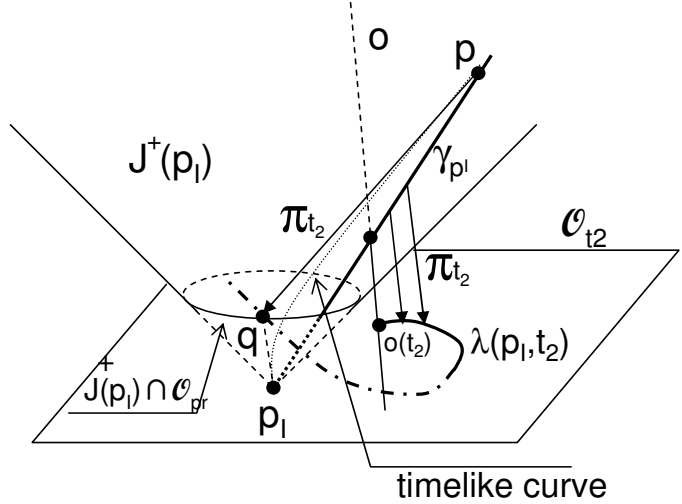}
\caption{The LRO of a point $p$, will 
not belong to the
causal future of $p$
itself. }\label{fig:fig6}
\end{figure}
We are now at the position to state the main theorem. %Now we will conclude the theorem of a first version.
\begin{theorem}[The half period theorem]
If the half period of the binary coalescence system elapses during $(t_1,t_2)$,
%Before the half period of the binary coalescence system elapses, 
%its 
two black holes merge 
into a %and become a
single black hole at a time $t'\in (t_1,t_2)$.
\end{theorem}
\begin{proof}
First note that the restriction $H\cap \OO$ of $H$ on $\OO$
is a null hypersurface in $\OO$ generated by null geodesics, each without a future end point.
By the definition of the half period of the binary coalescence system,
there is a point $p$ on $H_{\rm I}\cap \OO_{t_1}$ such that the trajectory of its LRO 
$\bigcup_{t\in(t_1,t_2)}\lambda(p,t)$ intersects with the orbit 
$\bigcup_{t\in(t_1,t_2)}\phi_t^I(p')$ of every point $p'$ on $H_{\rm I}$.
This implies that a null geodesic generator of $J^+(p)$,
which is also a null geodesic generator of $H_{\rm I}\cap\OO$ through $p$,
intersects with its own LRO $\lambda(p,t_p)$ at a time $t_p\in(t_1,t_2)$.
In the same way, there is a point $q$ on $H_{\rm II}\cap\OO_{t_1}$ such that
a null geodesic generator of $J^+(q)$ intersects with the LRO $\lambda(q,t_q)$
of $q$ at a time $t_q\in (t_1,t_2)$.
That is impossible is an immediate consequence of the Lemma~\ref{lem:1}.

\end{proof}

%\section{interpretation and discussion}
\section{Discussion}
First, we comment on the general covariance of our result.
The set up of the problem here might seem to depend on a specific coordinate system.
In fact, we prepare a time function and a timelike vector field to define the LRO.
This obviously corresponds to a specific choice of the time coordinate and $x^i={\rm const.}$
$(i=1,2,3)$ lines. 
Then, the LRO of a point $p$ is regarded as 
the line $x^i=x^i(s)$ on $t={\rm const.}$ surface
obtained by projecting points on the future directed null geodesic $x^\mu=x^\mu(s)$
$(\mu=0,1,2,3)$ starting from $p$
into the $t={\rm const.}$ surface along the $x^i={\rm const.}$ lines.
In this sense, the definition of the LRO, hence that of the half period
of the binary black hole system, depends on the coordinate system chosen.
Nevertheless, the half period theorem is formulated in a covariant manner
in the sense that
it holds for arbitrary choice of such ordinary coordinate system,
where ordinary coordinate we mean is such that a $t={\rm const.}$ surface is a spacelike
hypersurface and an $x^i={\rm const.}$ line is a timelike curve.

There are at least a couple of shortcomings in the definition of the LRO.
One is that the LRO of a point $p$ can be empty set when the congruence of the light rays
from $p$ to the orbit $o$ of the origin has caustics before reaching $o$.
Another is that the LRO might not be a sufficiently long curve, so that
it cannot be used as a goal line of the halfway around.

The statement of the half period theorem might sound extraordinary,
since it seems to contradict an existence of the quasi-stationary phase 
of binary black hole system. 
However, we would like to emphasis that it states an ordinary thing that
a comoving observer of a black hole does not exceed a speed of the light,
and that it does not contradict the quasi-stationary motion of the binary black holes.
Recently, there have appeared numerical results~\cite{NS1,NS2} showing that
binary black holes go around many times, which apparently contradicts our theorem.
In these numerical computations, apparent horizons, not the event horizons, are searched in the numerical space-time. 
There are two possibilities explaining this apparent contradiction.
The first one is the following. 
We expect that each apparent horizon is surrounded by the event horizon.
Then, there is a possibility that a pair of apparent horizons going around each other
is already enclosed by a single event horizon. In other words, while apparent horizons
go around many times, their event horizons quickly merge into one.
The second possibility is that the coordinate system used in the numerical simulation
 is superluminal, where $x^i={\rm const.}$ lines become space-like.

We expect that at least the second possibility is correct.
First, the apparent horizon, when regarded as a dynamical surface, become a spacelike
hypersurface in the space-time. This means that the apparent horizon has a confinement 
property. Therefore, each comoving observer, once entered the trapped region, never exit
unless the apparent horizon disappears. Hence, if the apparent horizons seem to go around
many times, the coordinate system describing this will be a superluminal one.

Finally, we speculate on implications of our theorem to astrophysical observation.
We cannot directly observe the event horizon. What are observed are light rays which
miss the event horizon. 
Let us consider the situation where we can directly observe shadows of a binary system
due to the existence of bright background light. 
This will be possible at least in principle.
If the binary system consists of ordinary dark stars other than black holes
and we observe it in the orbital surface (or with the maximal inclination angle),
we will see two shadows intersect many times each other.
While if it consist of black holes, we just observe two shadows merging into one exactly
once.

\vspace{25pt}
%%%%%%%%%%%%%%%%%%%%%%%%%%%%%%%%%%%%%%%%%%%%%%%%%%%%%

\end{document}